\documentclass[5p]{elsarticle}
\usepackage{lineno,hyperref}
\usepackage{xcolor}
\usepackage{amsmath}
\usepackage{amsfonts}
\usepackage{amssymb}
\usepackage{graphicx}
\usepackage{subfigure}
\journal{}

\begin{document}
\begin{frontmatter}
\title{Massive Argon Space Telescope (MAST): \\ a concept of heavy time projection chamber for $\gamma$-ray astronomy \\ in the 100 MeV --- 1 TeV energy range}
\author[Scobeltsyn,ICRR]{Timur Dzhatdoev\corref{mycorrespondingauthor}}
\cortext[mycorrespondingauthor]{Corresponding author}
\ead{timur1606@gmail.com}
\author[Scobeltsyn,MSU]{Egor Podlesnyi}
\address[Scobeltsyn]{Federal State Budget Educational Institution of Higher Education, M.V. Lomonosov Moscow State University, Skobeltsyn Institute of Nuclear Physics (SINP MSU), 1(2), Leninskie gory, GSP-1, 119991 Moscow, Russia}
\address[ICRR]{Institute for Cosmic Ray Research, University of Tokyo, 5-1-5 Kashiwanoha, Kashiwa, Japan}
\address[MSU]{Federal State Budget Educational Institution of Higher Education, M.V. Lomonosov Moscow State University, Department of Physics, 1(2), Leninskie gory, GSP-1, 119991 Moscow, Russia}

\begin{abstract}
We explore the concept of liquid Argon time projection chamber (TPC) for $\gamma$-ray astronomy in the 100 $MeV$ -- 1 $TeV$ energy range. We propose a basic layout for such a telescope called \textit{MAST}. Using a last-generation rocket such as \textit{Falcon Heavy}, it is possible to launch a detector with the effective area and the differential sensitivity about one order of magnitude better than the \textit{Fermi-LAT} ones. At the same time, the \textit{MAST} concept allows for an excellent angular resolution, 3--10 times better than the \textit{Fermi-LAT} one depending on the energy, and good energy resolution ($\approx$ 20 \% at 100 $MeV$ and 6--10 \% for the 10 $GeV$-- 1 $TeV$ energy range). We show that such a telescope would be instrumental in a broad range of long-standing astrophysical problems.
\end{abstract}
\begin{keyword}
primary $\gamma$-rays\sep space $\gamma$-ray telescopes\sep time projection chambers
\end{keyword}
\end{frontmatter}

\section{Introduction}

The Large Area Telescope (\textit{LAT}) \cite{Atwood2009} onboard the \textit{Fermi} mission have proved to be a great success of $\gamma$-ray astronomy. In particular, \textit{Fermi-LAT} detected about three thousand sources above 4$\sigma$ significance \cite{Acero2015}, put upper limits on the dark matter (DM) annihilation cross section \cite{Ackermann2015a} from observations of dwarf spheroidal galaxies, measured the extragalactic $\gamma$-ray background (EGRB) \cite{Abdo2010,Ackermann2015b} implying constraints on ultra-high energy cosmic ray and neutrino sources \cite{Berezinsky2011,Ahlers2010,Gelmini2012}, as well as on annihilating \cite{Ackermann2015c} and decaying \cite{Kalashev2016,Cohen2017} DM models, constrained the flux and spectrum of the extragalactic background light (EBL) \cite{Ackermann2012}. Additionally, \textit{Fermi-LAT} data may be used to put constraints on the extragalactic magnetic field (EGMF) parameters \cite{Neronov2010,Ackermann2018}.

Further progress in $\gamma$-ray astronomy would require a new instrument with improved sensitivity, angular and energy resolution, and wide energy coverage range. The fast development of astronautics and the reduction of prices on transportation services in space \cite{Falcon-Heavy} would soon allow for the launching of a very heavy $\gamma$-ray telescope with the mass up to 30--40 $t$. In this paper we propose a concept of a $\gamma$-ray telescope filled with pure liquid Argon and operating as time projection chamber (TPC) \cite{Nygren1974,Marx1978,Rubbia1977,Rubbia2011} called \textit{MAST} \footnote{not to be confused with Mega Amp Spherical Tokamak \cite{MAST} and other experiments and projects of the same abbreviation} (an abbreviation from ``Massive Argon Space Telescope'').

Many possible designs of TPC space-based $\gamma$-ray detectors filled with noble gases, sometimes in condensed phase, were considered before (e.g. \cite{Aprile1993,Aprile2008,Bernard2013,Bernard2013-Erratum,Hunter2014,Caliandro2014,Gros2018}). However, most of these were designed for polarimetric observations in the MeV-GeV energy range. In this paper we show, for the first time, that the idea to use a liquid noble gas TPC transcends the above-mentioned task, and is useful for a wider energy range, from 100 $MeV$ up to 1 $TeV$, and potentially even above 1 $TeV$. The present paper is organized as follows. In Sect.~\ref{sect:detector} we describe the basic geometry of the proposed instrument. In Sect.~\ref{sect:performance} we present the expected dependence of the effective area, angular and energy resolution vs. primary energy and calculate the point-like source differential sensitivity. After a brief discussion in Sect.~\ref{sect:discussion}, we conclude in Sect.~\ref{sect:conclusions}. All graphs in this paper were produced with the {\bf ROOT} software \cite{Brun1997}.

\section{The basic geometry of the \textit{MAST} telescope \label{sect:detector}}

\begin{figure}[tb]
\centering
\includegraphics[width=15pc]{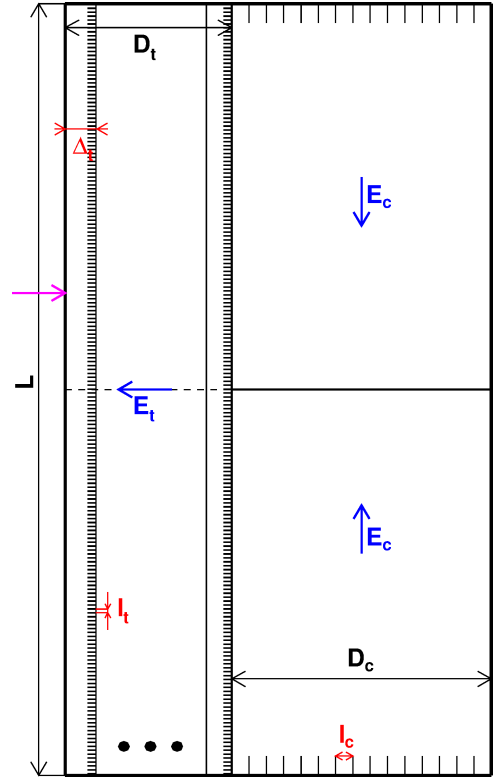}\hspace{2pc}%
\begin{minipage}[t]{16pc}\caption{A sketch of the proposed $\gamma$-ray detector \textit{MAST} (not to scale). See the text for more details.}
\label{Fig1}
\end{minipage}
\end{figure}

A simplified scheme of \textit{MAST} is shown in Fig.~\ref{Fig1}. The direction of a primary $\gamma$-ray, assuming normal incidence, is shown by magenta arrow. The detector with the overall dimensions\footnote{in what follows we assume the rectangular box shape for simplicity} $L\times L \times D$ with $L$= 400 $cm$ consists of the two main modules, namely, a tracker (thickness $D_{t} = 50$ $cm$) and a calorimeter (thickness $D_c = 110$ $cm$), so that $D=D_{t}+D_{c}$. Both modules are supposed to be filled with liquid argon (density $\rho$=1.4 $g/cm^{3}$ and radiation length $X_{0}$= 14 $cm$, see Tables 2.1 and 2.6 of \cite{Aprile2006}). The total sensitive mass of the \textit{MAST} telescope is $M= \rho\cdot L^{2}D$= 35.8 $t$, still within the capabilities of the \textit{Falcon Heavy} rocket \footnote{the maximal payload to the 185 $km\times$ 185 $km\times$ 28.5$^{\circ}$ low Earth orbit amounts to 63.8 $t$ \cite{Falcon-Heavy,SpaceLaunch}}. Assuming the orbit similar to that of the \textit{Fermi} mission (565 $km\times$ 565 $km\times$ 25.5$^{\circ}$ \cite{Atwood2009}), we estimate the maximal payload of the Falcon Heavy rocket for this orbit to be $\approx0.7\cdot63.8$ $t$ $\approx$ 44.7 $t$ \footnote{For the \textit{Fermi} mission, out of 4450 $kg$ total mass, only 3097 $kg$ is the payload, while 1002 $kg$ is reserved for a dry bus and 351 $kg$ for propellant \cite{FermiPayload}, hence the factor of 3097/4450$\approx$0.7}.

The tracker, in turn, consists of $N_{t}$=50 layers (only two of these layers are shown in Fig.~\ref{Fig1}). The thickness of each layer is $\Delta_t$=1 $cm$. A uniform electric field $E_{t}$= 3 $kV/cm$ normal to the layers permeates the tracker medium (denoted as horizontal blue arrow in Fig.~\ref{Fig1}). All layers of the tracker are equipped with a readout device with the longitudinal sampling $l_{t}$= 100 $\mu m$ in both dimensions collecting ionization electrons (two of these readouts are shown in Fig.~\ref{Fig1} by horizontal hatching).

The calorimeter is a single volume splitted by a cathode leaf and permeated by a uniform electric field $E_{c}$= 500 $V/cm$ (denoted as vertical blue arrow in Fig.~\ref{Fig1}) directed normally to the symmetry axis of the telescope (denoted as horizontal dashed line in Fig.~\ref{Fig1}). The calorimeter has two readouts on its walls with the longitudinal sampling $l_{c}$ = 1 $mm$ (shown in Fig.~\ref{Fig1} by vertical hatching).

\section{The performance of the \textit{MAST} telescope \label{sect:performance}}

Below we neglect the passive material of the detector (``dead material'') and consider only $\gamma$-ray events with conversion vertex in the first 36 layers of the tracker; thus, the effective thickness of the tracker is 36 cm=$D_{t}-X_{0}$. We also neglect saturation of the detector at high radiation energy densities. Additionally, following \cite{Bernard2013} we assume that the reconstruction is 100 \% effective. We consider only $\gamma$-ray conversion in the nuclear field. Events due to primary $\gamma$-ray conversion on atomic electrons (``triplet events'') may yield some additional information.

\subsection{Effective area}

\begin{figure*}[tb]
\begin{minipage}{0.47\textwidth}
\centering
\includegraphics[width=18pc]{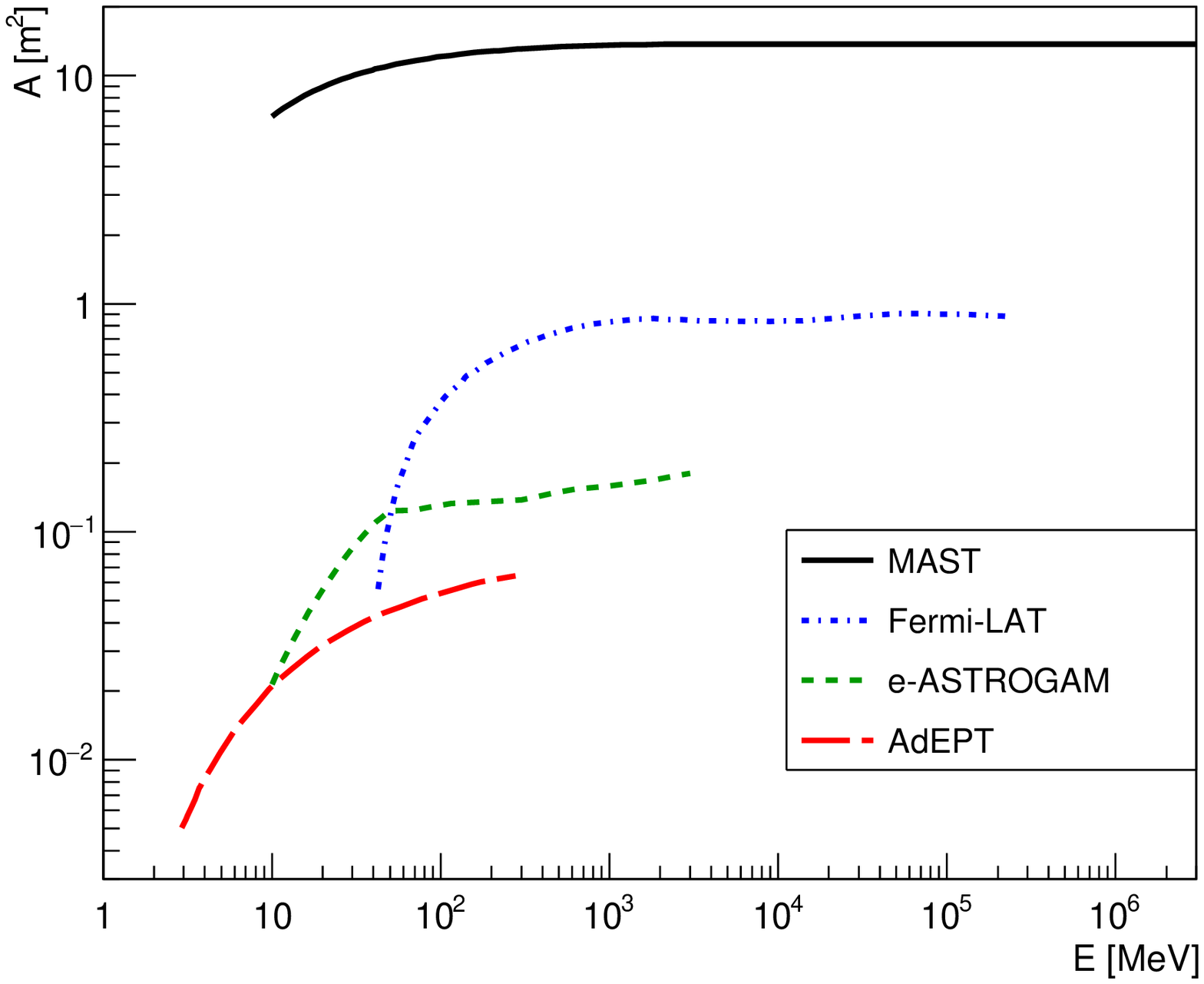}\hspace{2pc}%
\end{minipage}
\hfill
\begin{minipage}{0.47\textwidth}
\centering
\includegraphics[width=18pc]{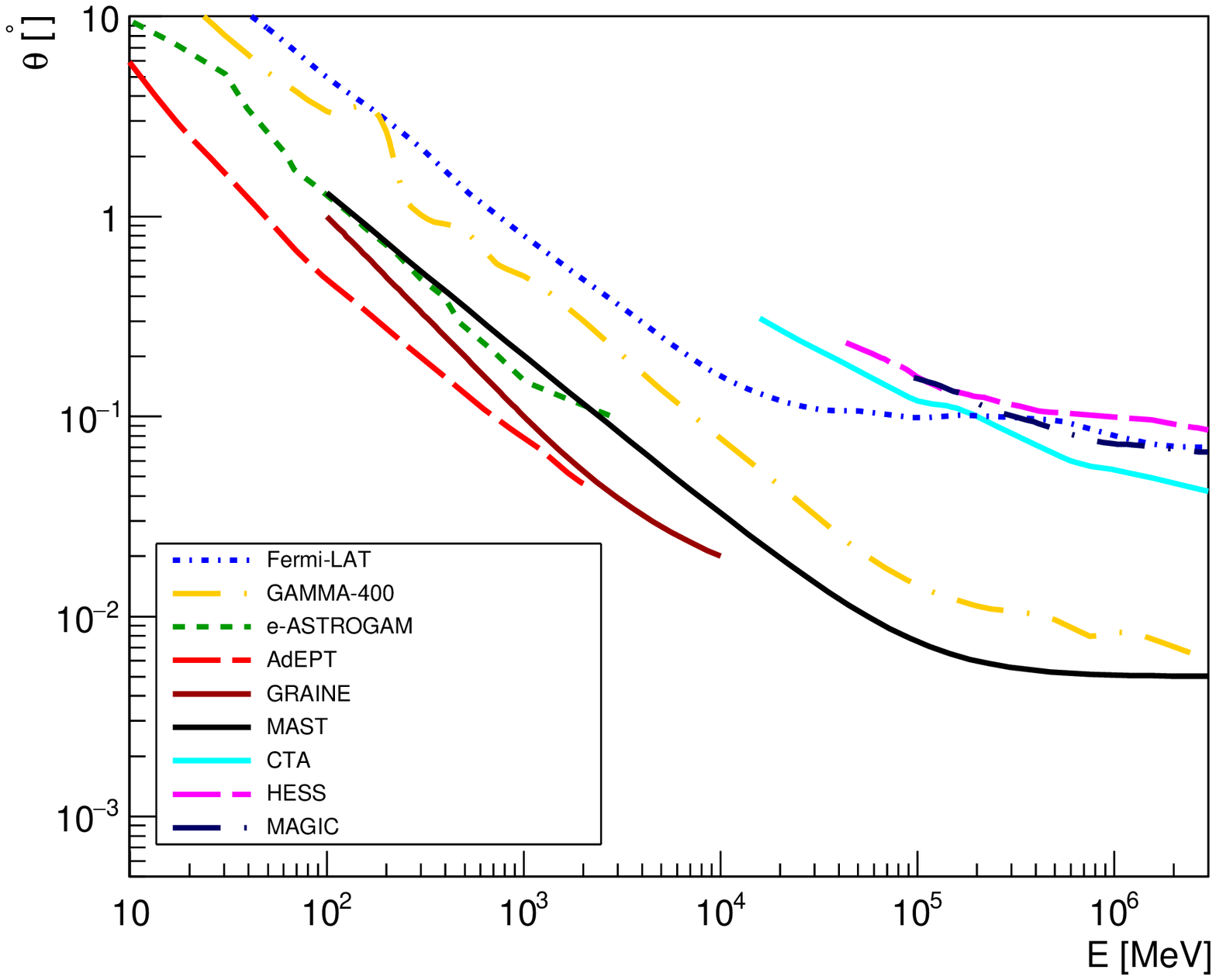}\hspace{2pc}%
\end{minipage}
\vspace{1.5pc}
\vfill
\begin{minipage}{0.47\textwidth}
\centering
\includegraphics[width=18pc]{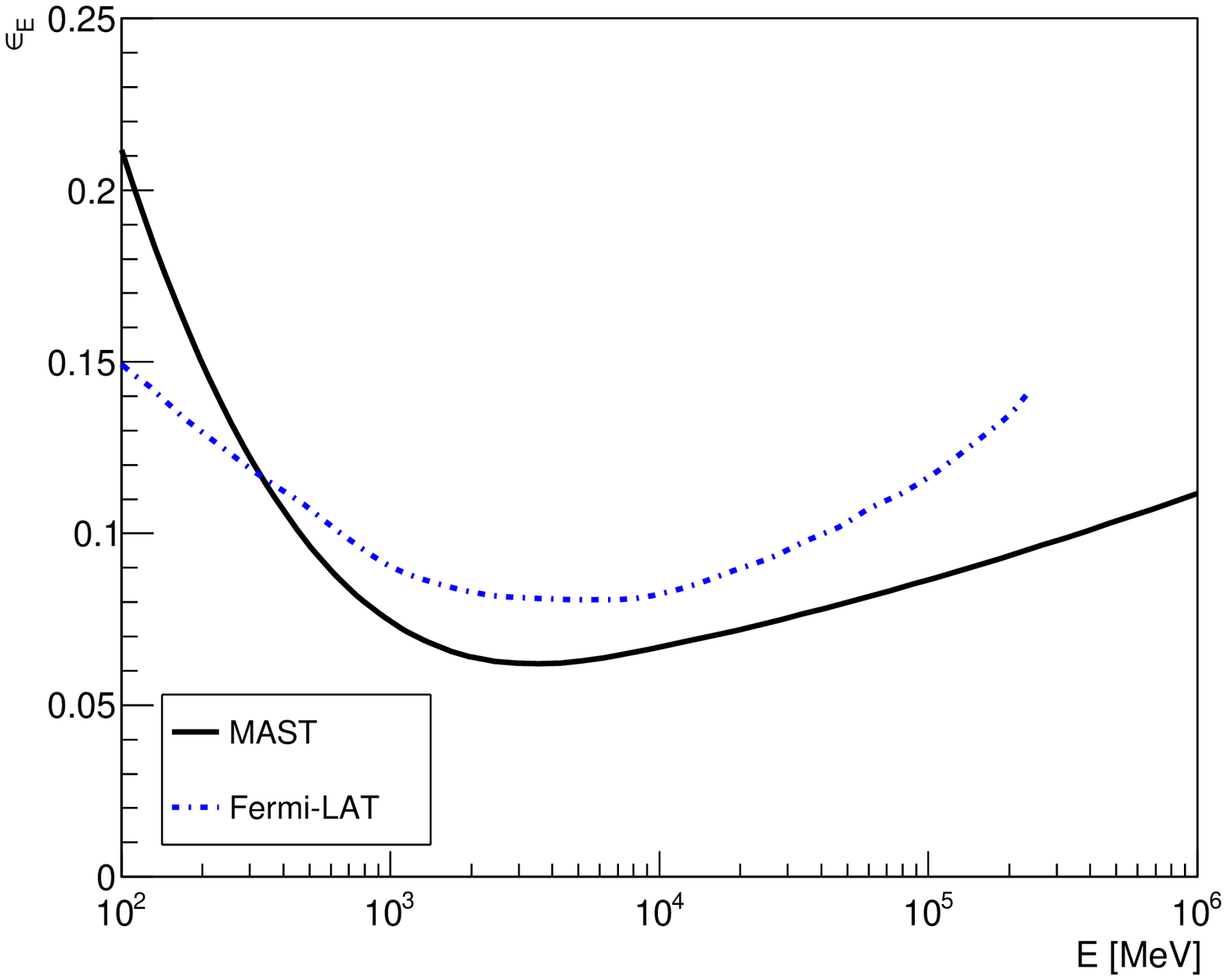}\hspace{2pc}%
\end{minipage}
\hfill
\begin{minipage}{0.47\textwidth}
\centering
\includegraphics[width=18pc]{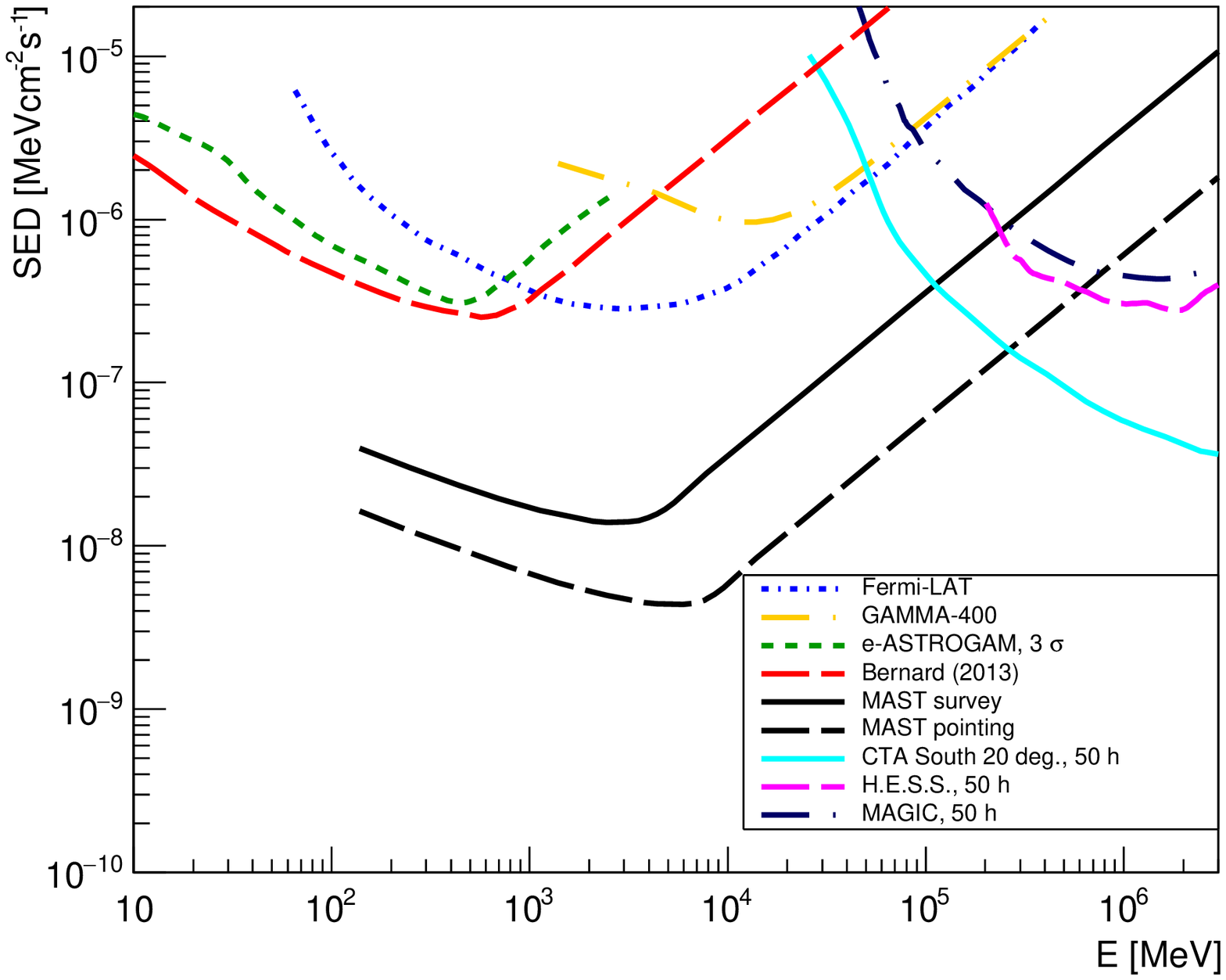}\hspace{2pc}%
\end{minipage}
\caption{Performance of the \textit{MAST} detector in comparison with other instruments. Top-left: effective area. Top-right: angular resolution. Bottom-left: energy resolution. Bottom-right: point-like source differential sensitivity.} \label{Fig2}
\end{figure*}

We estimate the effective area of the \textit{MAST} telescope for normal incidence using two equivalent approaches as follows: \\
\begin{equation}
\begin{split}
A(E)= L^{2}\left(1-exp\left[-\mu_{p}(E)\rho\left(D_{t}-X_{0}\right) \right]\right)= \\ L^{2}\left(1-exp\left[-\sigma_{p}(E)n\left(D_{t}-X_{0}\right) \right]\right),
\end{split}
\end{equation}
where $\mu_{p}$ $[cm^{2}/g]$ is the pair mass attenuation coefficient as a function of the primary $\gamma$-ray energy $E$, $\sigma_{p}$ $[cm^{2}]$ is the pair production cross section, and $n$ $[cm^{-3}]$ is the concentration of Argon atoms. The effective area of \textit{MAST} calculated following the latter approach with $\sigma_{p}$ according to {\bf Geant4} \cite{Agostinelli2003,Allison2006,Allison2016} Physics Reference Manual \cite{Geant4PRM} (version 10.4, Subsection 6.5.1, equation 6.4) \footnote{we assumed $\sigma_{p}=const$ at $E>$8.7 $GeV$} is presented in Fig.~\ref{Fig2} (top-left) in comparison with the same quantity for \textit{Fermi-LAT} \cite{Atwood2009} and for the projected $\gamma$-ray telescopes \textit{ADEPT} \cite{Hunter2014} and \textit{e-ASTROGAM} \cite{DeAngelis2018}. The relative difference between the values of $A$ calculated with the two above-mentioned approaches is less than 3 \% in the 10 $MeV$ -- 100 $GeV$ energy range.

\subsection{Angular resolution}

The single-photon angular resolution (68 \% containment radius) was estimated as follows: \\

\begin{equation}
\sigma_{\theta}=\sqrt{\sum_{i=1}^{4}\sigma_{\theta i}^{2}},
\end{equation}
where $\sigma_{\theta1}$ is the single-track angular resolution component due to finite detector spatial resolution, $\sigma_{\theta2}$ is the single-track angular resolution component due to multiple scattering, $\sigma_{\theta3}$ is the so-called ``kinematical limit'', and $\sigma_{\theta4}$ is the contribution from the uncertainty of the secondary electron and positron momentum absolute value measurement (for a discussion of pair production kinematics relevant for $\gamma$-ray telescopes see e.g. \cite{Gros2017,Bernard2018}).

Following \cite{Bernard2013,Bernard2013-Erratum},
\begin{equation}
\sigma_{\theta1}= \frac{2\sigma_{d}}{x}\sqrt{\frac{3}{N+3}}
\end{equation}
and
\begin{equation}
\sigma_{\theta2}= \frac{(2\sigma_{d})^{1/4}l_{t}^{1/8}}{X_{0}^{3/8}}\left(\frac{p_{0}}{p}\right)^{3/4},
\end{equation}
where $x$ is the tracking length (we assume $x=X_{0}$), $N=x/(3\sigma_d) \approx 560$ is the number of samplings \footnote{In the expression for $N$ we assume that the segments of the track become statistically independent at distances larger than $3\sigma_d$}, and $p_{0}$= 13.6 $MeV/c$. The spatial resolution $\sigma_{d}$ is limited mainly by the longitudinal sampling $l_{t}$ and the spatial extension of charge carrier cloud (which, in turn, is defined by the diffusion process, see eq. (3.4) of \cite{Aprile2006}):
\begin{equation}
\sigma_{d}= \sqrt{\frac{l_{t}^{2}}{12}+\frac{K_{D}\Delta_{t}}{v_{d}}}.
\end{equation}
Here $K_{D}$= 16 $[cm^{2}/s]$ \cite{Doke1981} (Fig. 10) is the perpendicular diffusion coefficient and $v_{d}$= 2.6$\cdot 10^{5}$ $[cm/s]$ \cite{Doke1981} (Fig. 7) is the charge carrier drift velocity (see also \cite{Miller1968,Shibamura1975,Shibamura1979,Yoshino1976} and \cite{Aprile2006} for discussion). If, instead, we consider only the segments of the track that are immediately adjacent to the readout device, then $N = x/\Delta_t = 14$ and $\sigma_{d}= l_{t}/\sqrt{12}$. Both options for the $(N,\sigma_{d})$ values result in comparable values of $\sigma_{\theta1}$.

The kinematical limit $\sigma_{\theta3}$ is due to the fact that the recoil nucleus is very hard to register. We assume that the recoil nucleus is not seen. The kinematical limit $\sigma_{\theta3}$ was calculated according to \cite{Jost1950} with correction of \cite{Borsellino1953} using tables for the Argon atomic form-factor presented in \cite{Hubbell1979}.

Finally, an uncertainty in electron ($p_{-}$) and positron ($p_{+}$) momenta absolute value measurement translates into some uncertainty of the reconstructed primary $\gamma$-ray direction. This contribution to the angular resolution ($\sigma_{\theta4}$) was estimated with a dedicated {\bf Geant4} (version 4.10.04.p02) application by calculating \textit{a change} \footnote {this comparison was made at the conversion vertex; calculating \textit{the change} in the reconstructed pair direction allowed us to single out the $\sigma_{\theta4}$ contribution} in the reconstructed pair direction $\vec p_{+}+\vec p_{-}$ assuming true MC momenta values in the first case and equal distribution of the momenta between the electron and the positron in the second case. 

The result of our calculation for the full angular resolution $\sigma_{\theta}$ is shown in Fig.~\ref{Fig2} (top-right) in comparison with the same quantity for other $\gamma$-ray detectors, namely \textit{Fermi-LAT} \cite{Atwood2009}, \textit{GAMMA-400} \cite{Galper2013, Galper2018, Topchiev2019}, \\ \textit{e-ASTROGAM} \cite{DeAngelis2018}, \textit{AdEPT} \cite{Hunter2014}, \textit{GRAINE} \cite{Takahashi2015}, as well as for imaging atmospheric Cherenkov telescopes (IACTs) \textit{CTA}, \textit{H.E.S.S.}, and \textit{MAGIC}. $\sigma_{\theta}$ for these IACTs was taken from \cite{CTA2018}; these experiments are described in \cite{Acharya2013,Bernlohr2013} (\textit{CTA}), \cite{Hinton2004,Bonnefoy2018} (\textit{H.E.S.S.}), and \cite{Aleksic2016a,Aleksic2016b} (\textit{MAGIC}). For all mentioned $\gamma$-ray instruments, we take the results on the angular resolution presented by the authors at face value. For VERITAS \cite{Krennrich2004,Park2016} the angular resolution is qualitatively similar; it is not shown to avoid confusion of the graph. A discussion of the point spread function of IACTs is available in \cite{DaVela2018}. We note that $\sigma_{\theta}$ for \textit{MAST} and \textit{e-ASTROGAM} is comparable notwithstanding the two-order difference in the effective area of these projected instruments.

\subsection{Energy resolution}

The energy resolution was estimated with the same {\bf Geant4} application that was used in the previous subsection. We simulated the development of electromagnetic cascades inside the model of the \textit{MAST} detector for five values of the primary $\gamma$-ray energy ($E$= 100 $MeV$, 1 $GeV$, 10 $GeV$, 100 $GeV$, 1 $TeV$) and recorded all the tracks of cascade electrons above the energy threshold of 10 $MeV$. The critical energy for Argon is $\approx33$ MeV for $e^{-}$ and $\approx32$ $MeV$ for $e^{+}$ \cite{PDG2018}.

The primary energy was reconstructed as follows: 1) using the integral over the cascade curve at $E\le$100 $GeV$ as the energy estimator and 2) by normalizing to the following analytical expression for the cascade curve (see eq. (33.36) in \cite{Tanabashi2018}) at $E$= 1 $TeV$:
\begin{equation}
\frac{dE}{dt}= C_{0}b \frac{(bt)^{a-1}e^{-bt}}{\Gamma(a)},
\end{equation}
where $a$, $b$, and $C_{0}$ are the free parameters of the fit, $t$ is the thickness in radiation lengths, and $\Gamma$ denotes the gamma function. For the case of $E$= 100 $GeV$ and the first energy estimation method we introduced an additional correction to the estimated value of the primary energy depending on the depth of the shower's maximum. Using a simple linear correction, we were able to improve the resulting relative energy resolution $\epsilon_{E}$ from 11.4 \% to 8.6 \%. For the case of $E$= 1 $TeV$ we excluded $\approx$3 \% of cascades which revealed fitting problems.

The dependence of $\epsilon_{E}$ on $E$ is shown in Fig. 2 (bottom-left) in comparison with the \textit{Fermi-LAT} energy resolution \cite{Atwood2009}. We note that it appears possible to estimate the primary energy with a reasonable accuracy both at the low-energy region ($E$= 100 $MeV$--1 $GeV$), as well as at the high-energy region ($E$= 100 $GeV$--1 $TeV$), notwithstanding a relatively high critical energy in Argon (compared to the tungsten calorimeter of \textit{Fermi-LAT}) and a relatively low depth of the \textit{MAST} calorimeter that amounts to 11.4 radiation lengths. The details of this primary energy estimation procedure will be published elsewhere. Moreover, the results of \cite{Ankowski2010} indicate that a several-fold improvement in the energy resolution is achievable in the energy range of 0.5--5 $GeV$.

\subsection{Differential sensitivity}

Finally, we estimated the differential sensitivity of the \textit{MAST} telescope for point-like sources following the approach of \cite{Bernard2013} and assuming four bins per decade of energy, the threshold significance of 5 $\sigma$, and the effective area and angular resolution as described above. We also require at least ten events in every energy bin. The EGRB with the spectrum in the form $dN/dE= C_{0}(E/E_{0})^{-\gamma}exp(-E/E_{c})$ with $(C_{0}$= 7.8$\cdot 10^{-8}$ $MeV^{-1}cm^{-2}s^{-1}sr^{-1}$, $E_{0}$= 100 $MeV$, $\gamma$= 2.26, and $E_{c}$= 233 $GeV)$ (model C of \cite{Ackermann2015b}) was assumed as the background in this calculation. The resulting differential sensitivity of \textit{MAST} for ten years of observation is shown in Fig. 2 (bottom-right) in comparison with the same quantity for \textit{Fermi-LAT} \cite{Atwood2009}, \textit{GAMMA-400} \footnote{the table for the differential sensitivity was kindly provided by the \textit{GAMMA-400} team}, \textit{e-ASTROGAM} \cite{DeAngelis2018} (for the threshold significance of 3 $\sigma$), as well as for the projected TPC $\gamma$-ray telescope \cite{Bernard2013,Bernard2013-Erratum}, \textit{H.E.S.S.}, \textit{MAGIC} and \textit{CTA} (data for these IACTs are from \cite{CTA2018}). The sensitivity of IACTs is for 50 hours of observation. The sensitivity for \textit{MAST} is shown for two options: 1) ``the survey mode'' when the source is seen only a fraction of time $\eta$=~0.17 (solid curve) and 2) ``the pointing mode'' with $\eta$=~1 (dashed curve), and for \textit{GAMMA-400} --- for ``the pointing mode'' and ten years of observation. We note that a preliminary version of this result was presented in \cite{Dzhatdoev2018a} for different detector parameters.

\section{Discussion \label{sect:discussion}}

Calculations made in the previous section show that the \textit{MAST} $\gamma$-ray telescope could achieve a significant improvement over presently operating and projected space-based $\gamma$-ray detectors in terms of the effective area and differential sensitivity, at the same time retaining excellent angular resolution and reasonable energy resolution. The tracker and calorimeter of the \textit{MAST} detector should be complemented with a segmented anticoincidence detector (ACD) in order to suppress the background from charged particles. The operation of the ACD is possible thanks to the design of the MAST tracker with minimal ($\sim10^{2}-10^{3}$ $ns$) drift times for track segments immediately adjacent to the readout device.

In what follows we assume the trigger condition \\ $(S_{ACD}=0)AND(E_{dep}>30 MeV)$, i.e. that the signal in the ACD is below the threshold and the energy deposit inside the tracker exceeds 30 $MeV$. This condition allows to register 100 $MeV$ $\gamma$-rays without any loss in efficiency provided that the conditions specified in Sect. 2 are satisfied. Furthermore, assuming the background fluxes according to Fig. 12 of \cite{Atwood2009}, and the ACD with the efficiency $\alpha= 1-\delta$ with $\delta$= $3\cdot10^{-4}$ \cite{Moiseev2007}, we estimate the total residual rate for charged particles $R_{ch}\approx$30 $Hz$ (for comparison, the expected signal rate is $\sim$20 $Hz$).

For terrestrial $\gamma$-rays $R_{\gamma}\sim$500 $Hz$; for neutrons, using a separate {\bf Geant4} application and assuming the QGSP\_BERT interaction model, we obtained a similar value, $R_{n}\sim$500 $Hz$. These conservative estimates were obtained with a rough account of the angular resolution of terrestrial $\gamma$-rays and neutrons that is peaked towards the Earth while the telescope is oriented in the opposite direction. The total background rate $R_{b-Tot}\sim$1 $kHz$ is comparable with the downlink frequency $\sim$400 $Hz$ \cite{Atwood2009}. A more detailed analysis would result in lower $R_{b-Tot}$. To conclude, terrestrial $\gamma$-rays and neutrons represent a dangerous source of background, but still $R_{b-Tot}$ is low enough to allow data acquisition and avoid the overlapping of different events.

We note that the huge (total mass 68 $kt$) liquid Argon detectors of the DUNE project \cite{Acciarri2015} are presently in development. The successful operation of the \textit{ICARUS T 600} experiment \cite{Rubbia2011} once more proves that the liquid Argon TPC is an extremely cost-effective technique. Therefore, we estimate the capital costs of the \textit{MAST} and \textit{Fermi-LAT} detectors to be comparable notwithstanding the order-of-magnitude difference in their masses. The price of the \textit{Falcon Heavy} rocket 
launch \cite{Falcon-Heavy} is only twice the price of the \textit{Delta-II} rocket \cite{Delta-2} which was used to carry the \textit{Fermi-LAT} telescope. Future progress in rocket technologies could result in comparable capital costs of the \textit{MAST} and \textit{Fermi-LAT} experiments.

Possible astrophysical tasks for the \textit{MAST} instrument include, but are not confined to the EBL  \cite{Stecker1993,DeJager1994,Ackermann2012,Fermi-LAT2018} and EGMF \cite{Neronov2010,Ackermann2018} measurement (for a review see \cite{Durrer2013,Han2017}), the study of the $\gamma$-ray -- neutrino connection in blazars \cite{Stecker1991,Nellen1993,Kalashev2015,IceCube2018a,IceCube2018b}, observations of neutron star mergers \cite{Kasliwal2017,Abbott2017}, the search for $\gamma$-rays from black hole mergers \cite{Abbott2016,Connaughton2016}, $\gamma \rightarrow ALP$ oscillation \cite{Raffelt1988} in active galactic nuclei spectra \cite{Horns2012,Ajello2016,Kartavtsev2017,Montanino2017,Galanti2018a,Galanti2018b,Korochkin2018}, Lorentz invariance violation search \cite{Coleman1999,Kifune1999,Abdalla2018} (see, however, \cite{Rubtsov2017}), and DM annihilation or decay signatures search \cite{Ackermann2015c,Kalashev2016,Cohen2017}. Detailed investigation of the \textit{MAST} sensitivity to these processes is beyond the scope of the present paper. Preliminary estimates obtained with intergalactic cascade calculation techniques developed in \cite{Dzhatdoev2017a,Dzhatdoev2017b,Fitoussi2017} show a good sensitivity to the EGMF strength down to 10 $aG$ and coherence length of 1 $kpc$ (see also \cite{Meyer2016,Dzhatdoev2018b}). The results of this study will be published elsewhere. Finally, we note that \textit{MAST} could represent an excellent low-energy counterpart for ground-based $\gamma$-ray detectors such as CTA \cite{Acharya2013} and LHAASO \cite{Cui2014,Tian2018}.

\section{Conclusions \label{sect:conclusions}}

In this paper we considered a concept of a $\gamma$-ray telescope called \textit{MAST} based on heavy time projection chamber (TPC) filled with liquid Argon. Estimates of the effective area, angular resolution, energy resolution, and differential sensitivity show a great potential of such an instrument in a wide range of $\gamma$-astronomical tasks. We conclude that constructing and operating the \textit{MAST} detector could facilitate a significant advance in astroparticle physics.

\section*{Acknowledgements}

This work was supported by the Russian Science Foundation (RSF) (project No 18-72-00083). We are greatly indebted to Dr. D. Bernard for very helpful discussions. We are grateful to Dr. S. Takahashi and Dr. N.P. Topchiev for providing us with the tables characterising the angular resolution of the \textit{GRAINE} instrument and the angular resolution and differential sensitivity of the \textit{GAMMA-400} telescope. We acknowledge helpful discussions with Dr. D.V. Chernov, I.A. Kudryashov, Prof. I.V. Moskalenko, Prof. K. Murase, Dr. G.I. Rubtsov, and Dr. A.V. Uryson.

\section*{References}
\bibliography{TPC-Concept.bib}
\end{document}